\documentstyle[twoside,fleqn,epsf,espcrc2]{article}

\newcommand{\AmS}{{\protect\the\textfont2
  A\kern-.1667em\lower.5ex\hbox{M}\kern-.125emS}}

 \newcommand{\postbb}[3]
 {\setlength{\epsfxsize}{#3\hsize}
  \centerline{\epsfbox[#1]{#2}}}

\newcommand{\skipblk}[1]{}                                                      
\newcommand{\beqa}{\begin{eqnarray}}
\newcommand{\eeqa}{\end{eqnarray}} 
\newcommand{\bqa}{\begin{eqnarray}}
\newcommand{\eqa}{\end{eqnarray}}

\newcommand{\doub}[3]{\mbox{$                                                   
\left( \begin{array}{c} #1    \\ #2  \end{array} \right)_#3$}}                  

\newcommand{\PR}[3]{Phys. Rev. {\bf #1}, #2 (19#3)}                             
\newcommand{\PL}[3]{Phys. Lett. {\bf #1}, #2 (19#3)}                            
\newcommand{\NP}[3]{Nucl. Phys. {\bf #1}, #2 (19#3)}                            
\newcommand{\PRL}[3]{Phys. Rev. Lett. {\bf #1}, #2 (19#3)}

\newcommand{\eg}{{\em e.g., }}                                                  
\newcommand{\ie}{{\em i.e., }}

\newcommand{\x}{\mbox{$\times$}}

\newcommand{\beq}{\begin{equation}}                                             
\newcommand{\eeq}{\end{equation}}

\newcommand{\RA}{\mbox{$\rightarrow$}}

\def\mxth{\mathsurround=0pt }
\def\xversim#1#2{\lower2.pt\vbox{\baselineskip0pt \lineskip-.5pt
  \ialign{$\mxth#1\hfil##\hfil$\crcr#2\crcr\sim\crcr}}}             
\def\simgr{\mathrel{\mathpalette\xversim >}}

\newcommand{\mstr}{\mbox{$M_{\rm str}$}}                              
 
\title{Implications of Solar and Atmospheric Neutrinos}

\author{Paul Langacker\address{Department of Physics and Astronomy \\ 
          University of Pennsylvania, Philadelphia PA 19104-6396, USA}}
       
\begin{document}

\begin{abstract}
The importance of non-zero neutrino mass
as a probe of particle physics, astrophysics, and cosmology is
emphasized. The present status and future prospects for the solar
and atmospheric neutrinos are reviewed, and the implications for
neutrino mass and mixing in 2, 3, and 4-neutrino schemes are
discussed. The possibilities for significant mixing between ordinary
and light sterile neutrinos are described.\end{abstract}

\maketitle

\section{NEUTRINO MASS}

Neutrino mass and properties are  superb simultaneous probes
of particle and astrophysics: 

\begin{itemize}

\item Decays and scattering processes involving neutrinos have
been powerful probes of the existence and properties of quarks,
tests of QCD, of the standard electroweak model and its parameters, 
and of possible
TeV-scale physics.
 
\item 

 Fermion masses in general are one of the major mysteries/problems
of the standard model.  Observation or nonobservation of the
neutrino masses  introduces a useful new perspective on the subject.

\item

Nonzero $\nu$ masses are predicted in most extensions of the standard
model.  They therefore constitute a powerful window on new physics at the
TeV scale, intermediate scales (e.g., $10^{12}$ GeV), or
the Planck scale.

\item

There may be a hot dark matter component to the universe.  If so,
neutrinos would be (one of) the most important things in the universe.

\item

The neutrino masses must be understood to fully exploit neutrinos
as a probe of the Solar core, of supernova dynamics, and of
nucleosynthesis in the big bang, in stars, and in supernovae.

\end{itemize}


\section{THEORY OF NEUTRINO MASS} 
\label{theorysection}
There are a confusing variety of models of neutrino mass. Here,
I give a brief survey of the principle classes and of some of the terminology. 
For more
detail,  see~\cite{lr0}.

A Weyl two-component spinor is a left ($L$)-handed\footnote{The subscripts
$L$ and $R$ really refer to the left and right chiral projections. In
the limit of zero mass these correspond to left and right helicity
states.} \ particle state, $\psi_L$, which is necessarily associated by CPT
with a right ($R$)-handed antiparticle state\footnote{Which is referred
to as the particle or the antiparticle is a matter of convenience.}
\ $\psi^c_R$. One refers to active (or ordinary) neutrinos
as left-handed neutrinos
which transform as $SU(2)$ doublets with a charged lepton
partner. They therefore have normal weak interactions, as do their right-handed
anti-lepton partners, 
\beq \doub{\nu_e}{e^-}{L} \stackrel{\rm  CPT}{\longleftrightarrow}
\doub{e^+}{\nu^c_e}{R}. \eeq
Sterile\footnote{Sterile neutrinos are often referred to as
``right-handed'' neutrinos, but that terminology is
confusing and inappropriate when Majorana masses are present.}
\ neutrinos are  $SU(2)$-singlet neutrinos, which can be added to the 
standard model and  are predicted in most extensions. They have
no ordinary weak interactions except those induced by mixing with active
neutrinos. It is usually convenient to define the $R$ state as the
particle and the related $L$ anti-state as the antiparticle.
\beq N_R \stackrel{\rm  CPT}{\longleftrightarrow} N^c_L. \eeq
(Sterile neutrinos will sometimes also be denoted $\nu_s$.)

Mass terms describe transitions between right ($R$)
and left ($L$)-handed states.
A Dirac mass term, which conserves lepton number, involves transitions
between two distinct Weyl neutrinos
$\nu_L$ and $N_R$:
\beq
- L_{\rm Dirac}  =  m_D (\bar{\nu}_L N_R +
\bar{N}_R \nu_L)
  =  m_D \bar{\nu} \nu, \eeq
where the Dirac field is defined as $\nu \equiv \nu_L + N_R$.  Thus a
Dirac neutrino has four components $ \nu_L, \; \nu_R^c, \; N_R, \;
N_L^c$,
and the mass term allows a conserved lepton number $L = L_\nu
+ L_N$.  This and other types of mass terms can easily be generalized
to three or more families, in which case the masses become matrices.
The charged current transitions then involve
a leptonic mixing matrix (analogous to the
Cabibbo-Kobayashi-Maskawa (CKM) quark mixing  matrix), which
can lead to neutrino oscillations between the light neutrinos.

For an ordinary Dirac neutrino
the $\nu_L$ is active 
and the $N_R$ is
sterile.
The transition is $\Delta I = \frac{1}{2}$,
where $I$ is the weak isospin.  The mass requires $SU(2)$ breaking and
is generated by a Yukawa coupling
\beq  -L_{\rm Yukawa} = h_\nu (\bar{\nu}_e \bar{e})_L
\left( \begin{array}{c} \varphi^0 \\ \varphi^- \end{array} \right)
  N_R + H.C. \eeq
One has $m_D = h_\nu v/\sqrt{2}$,
where the vacuum expectation value (VEV) of
the Higgs doublet is
$ v = \sqrt{2} \langle \varphi^o \rangle = ( \sqrt{2} G_F)^{-1/2} =
246$ GeV,
and $h_\nu$ is the Yukawa coupling.
A Dirac mass is just like the quark and charged lepton masses, but
that leads to the question of    why it is so small: one  requires
$h_{\nu_e} < 10^{-10}$  to have $m_{\nu_e} < 10$ eV.  

A Majorana mass, which violates lepton number by two units $(\Delta L
= \pm 2)$, makes use of the right-handed antineutrino, 
$\nu^c_R$, rather than a separate Weyl neutrino.  It is a transition
from an antineutrino into a neutrino. Equivalently, it can be viewed
as the creation  or annihilation of two neutrinos, and if present
it can therefore lead to neutrinoless double beta decay.
The form of a Majorana mass term is
\beqa - L_{\rm Majorana}  & = &\frac{1}{2} m_T (\bar{\nu}_L \nu_R^c +
\bar{\nu}^c_R \nu_L )   = \frac{1}{2} m_T \bar{\nu} \nu \nonumber \\
& = & \frac{1}{2} m_T (\bar{\nu}_L C \bar{\nu}_L^T + H.C.),
 \eeqa
where $\nu = \nu_L +\nu^c_R$ is a self-conjugate two-component state
satisfying $\nu = \nu^c = C \bar{\nu}^T$, where $C$ is the
charge conjugation matrix.  If $\nu_L$ is active then
$\Delta I = 1$ and $m_T$ must be generated by either an elementary Higgs
triplet or by an effective operator involving two Higgs doublets
arranged to transform as a triplet. 

One can also have a Majorana
mass term
\beq - L_{\rm Majorana} = \frac{1}{2} m_N (\bar{N}^c_L N_R +
\bar{N}_R N^c_L ) \eeq
for a sterile neutrino. This has $\Delta I = 0$ and thus
can be generated by the VEV of a
Higgs  singlet\footnote{In principle this could also be generated by a bare
mass, but this is usually forbidden by higher symmetries in extensions 
of the standard model.}.

Some of the principle classes of models for neutrino mass are:
\begin{itemize}
\item A triplet majorana mass $m_T$ can be generated by
the VEV $v_T$ of a Higgs triplet field. Then, $m_T = 
h_T v_T$, where $h_T$ is the relevant Yukawa
coupling. Small values of $m_T$ could be due to a small scale
$v_T$, although that introduces a new hierarchy problem.
The simplest implementation
is the Gelmini-Roncadelli (GR) model~\cite{lr19},
in which lepton number is spontaneously broken by $v_T$. The
original GR model is now excluded by the LEP data on the $Z$
width.

\item
A very different class of models are those in which the neutrino
masses are zero at the tree level (typically because no sterile neutrino
or elementary Higgs triplets are introduced), but only generated by
loops \cite{lr57}, \ie
radiative generation.  Such models
generally require
the {\em ad hoc} introduction of new scalar particles at the TeV scale
with nonstandard
electroweak quantum numbers and lepton number-violating couplings.
They have also been introduced in an attempt to generate large electric or
magnetic dipole moments. They also occur in some  supersymmetric models with 
cubic $R$
parity violating terms in the superpotential~\cite{susynu}.

\item In the seesaw models~\cite{lr16}, a small Majorana mass
is induced by mixing between an active neutrino and a very heavy
Majorana sterile neutrino $M_N$. The light (essentially active)
state has a naturally small mass
\beq m_\nu \sim \frac{m_D^2}{M_N} \ll m_D. \eeq
There are literally hundreds of seesaw models, which differ in the scale
$M_N$ for the heavy neutrino (ranging from the TeV scale to grand unification
scale), the Dirac mass $m_D$ which connects the ordinary and sterile
states and induces the mixing (e.g., $m_D \sim m_u$ in most grand unified theory
(GUT) models, or $\sim m_e$ in left-right symmetric models), the patterns
of $m_D$ and $M_N$ in three family generalizations, etc. One can also
have mixings with heavy neutralinos in supersymmetric models with $R$ parity
breaking~\cite{susynu}, induced either by bilinears connecting Higgs and lepton
doublets in the superpotential or by the expectation values of scalar neutrinos. 

\item Superstring models often predict the existence of higher-dimensional
(nonrenormalizable) operators (NRO) such as
\beq -L_{\rm eff} = \bar{\psi}_L H \left(\frac{S}{\mstr}\right)^P \psi_R
+ H.C., \eeq
where $H$ is the ordinary Higgs doublet, $S$ is a new scalar field which is
a singlet under the standard model gauge group, and $\mstr \sim
10^{18}$ GeV is the string scale. In many cases $S$ will acquire
an intermediate scale VEV (e.g., $10^{12}$ GeV), leading to an
effective Yukawa coupling 
\beq h_{\rm eff} \sim v \left(\frac{\langle S \rangle}{\mstr}\right)^P
\ll v. \eeq
Depending on the dimensions $P$ of the various operators and on
the scale $\langle S \rangle$, it may be possible to generate
an interesting hierarchy for the quark and charged lepton masses
and to obtain naturally small Dirac neutrino masses~\cite{nro}.
Similarly, one may obtain triplet and singlet Majorana neutrino
masses, $m_T$ and $m_N$ by analogous higher-dimensional operators.
The former are small. Depending on the operators~\cite{nro} 
the latter may be either
small, leading to the possibility of significant mixing between ordinary
and sterile neutrinos~\cite{sterile}, or
large, allowing a conventional seesaw.

\item Mixed models, in which both Majorana
and Dirac mass terms are present, will be further discussed
in the section on sterile neutrinos.

\end{itemize}

\section{SOLAR NEUTRINOS}
Tremendous progress has been made recently in solar 
neutrinos~\cite{neuastro}.
For many years there was only one experiment, while now
there are a number that are running or finished, 
and more are coming on line soon.
The original goal of using the solar neutrinos to study the
properties of the solar core underwent a 30 year digression
on the study of the properties of the neutrino itself.
The quality of the experiments themselves and of related
efforts on helioseismology, nuclear cross sections, and solar modeling
is such
that the revised goal of simultaneously studying the
properties of the Sun and of the neutrinos is feasible. 

\subsection{Experiments}
The experimental situation is very promising.
We now have available the results of five
experiments, Homestake (chlorine)~\cite{homestake}, 
Kamiokande~\cite{superk}, GALLEX~\cite{gallex}, SAGE~\cite{sage}, and
Superkamiokande~\cite{superk}. Especially impressive are the successful 
$^{51}Cr$ source
experiments for  SAGE and GALLEX
 (which probe a combination of the extraction efficiencies
and the neutrino absorption cross section, yielding
$0.95 \pm 0.07^{+0.04}_{-0.03}$ of the expected rate), 
and the
successful
$^{71}As$ spiking experiment completed at the 
end of the GALLEX run to test the extraction
efficiency (yielding $R = 1.00 \pm 0.01$ for the ratio of 
actual to expected extractions).

Coming soon, there should be results from SNO,
Borexino, The Gallium Neutrino Observatory (GNO), and the next
phase of SAGE, which will yield much more detailed, precise,
or model independent
information on the $^8B$  (SNO~\cite{sno}), $^7Be$ 
(Borexino~\cite{borexino}), and
$pp$ (GNO, SAGE) neutrinos. Future generations of
even more precise experiments should especially be sensitive to the
$^7Be$ and $pp$ neutrinos~\cite{futuresolar}.
The overall goal of the program should be very ambitious, \ie \
to measure the arriving flux of $\nu_e, \nu_{\mu + \tau}$, and
$\nu_s$ (sterile neutrinos), and even possible antineutrinos,
for each of the initial flux components, as well
as to measure or constrain possible spectral distortions, day-night (earth) effects,
seasonal and solar cycle variations, and mixed  (\eg \
simultaneous spectral and day-night) effects.

\subsection{Interpretation}
The observed fluxes are in strong disagreement with the
predictions of the standard solar model (SSM). 
The overall rates are compared with the predictions of
the new Bahcall-Pinsonneault 1998 (BP 98) model~\cite{BP}
 in
 Table \ref{solarexp}, where it is seen that all of the fluxes
are much lower than the expectations. BP 98 contains
a number of refinements compared to earlier theoretical calculations,
but the most important changes are a 20\% (1.3 $\sigma$)
lower $^8B$ flux, as described below, and 1.1 $\sigma$ decreases
in the $^{37}Cl$ and $^{71}Ga$ capture rates. 
\begin{table*}[hbt]
\setlength{\tabcolsep}{1.5pc}
\newlength{\digitwidth} \settowidth{\digitwidth}{\rm 0}
\catcode`?=\active \def?{\kern\digitwidth}
\caption{Results of Solar neutrino experiments, compared
with the predictions of BP 98. The chlorine and gallium results
are in units of SNU ($10^{-36} s^{-1}$ captures per target atom),
and the water Cerenkov results are in units of $10^6/cm^2 s$.}
\label{solarexp}
\begin{tabular*}{\textwidth}{@{}l@{\extracolsep{\fill}}lll}
\hline
 & experiment  & BP-98 \\
\hline
Homestake    & $2.56 \pm 0.23$ &  $7.7^{+1.2}_{-1.0} $ \\
(chlorine)  &  & \\
GALLEX, SAGE    & $72.2 \pm 5.6$ &  $129^{+8}_{-6} $ \\
(gallium)          &  &\\
Kamiokande, SuperK          & $2.44 \pm 0.10$ & $5.15 \x (1^{+0.19}_{-0.14})$ \\
($\nu e \RA \nu e$) &        &    \\
\hline
\end{tabular*}
\end{table*}

Recent results in helioseismology~\cite{BP,cast,degl,helio,bahcall}
 leave little room for deviations
from the standard solar model. The eigen-frequencies effectively
measure the sound speed $T/\mu$, where $T$ and $\mu$ are respectively 
the temperature
and density, as a function of radial position, down to 5\% of the solar radius.
The results agree with the predictions of BP 98 to $\sim 10^{-3}$,
even though $T$ and $\mu$ individually vary by large values over
the radius of the Sun. This leaves very little room for
non-standard solar models (NSSM), which would typically have
to deviate by several percent to have much impact on the
neutrino flux predictions. The only aspect of the SSM relevant to the neutrino
fluxes that is not severely constrained are
nuclear cross sections, especially $S_{17}$ and $S_{34}$, which are
respectively proportional to the cross sections for $^8B$ and
$^7Be$ production, and to the absorption cross sections for the
radiochemical experiments.

The experimental and theoretical status of the
nuclear cross sections were critically
examined at a workshop at the Institute for Nuclear Theory in 1997 
(INT 97)~\cite{bahcall,INT}.
The participants recommended a lower $S_{17}$, by relying on the
best documented individual measurements rather than an average, and
also a larger uncertainty in $S_{34}$, both of which were incorporated
in BP 98. Haxton has recently argued~\cite{haxton}
 that there are still considerable
uncertainties in the $Ga$ absorption cross sections, but this possibility
is strongly disfavored by the $^{51}Cr$ source and $^{71}As$ spiking
experiments.

Even the relatively large
shift in $S_{17}$ advocated by INT 97 and used by BP 98
does little to change the basic
disagreement between the observations and the standard solar model.
Even if a particular NSSM could be consistent with helioseimology,
it would be difficult to account for the observations. The Kamiokande
and Superkamiokande results can be regarded (in the absence of
neutrino oscillations) as a measurement of the $^8B$ flux.
Subtracting this ``experimental'' $^8B$ flux from either the
gallium or chlorine predictions, the observed fluxes are still
inconsistent with the observed solar luminosity.

This line of reasoning is developed in the ``model-independent''
analyses of the neutrino flux components~\cite{analyses,hl,bks}, 
which can be viewed
as a measurement of ``global'' spectral distortions. The idea is
that all plausible astrophysical or nuclear physics modifications
of the standard solar model do not significantly distort the spectral
shape of the $pp$ or $^8B$ neutrinos: all that they can do is modify
the overall magnitude of the $pp$, $^7Be$, $^8B$, and minor flux
components. Furthermore, the observed solar luminosity places
a linear constraint on the $pp$, $^7Be$, and CNO fluxes (provided
that the time scale for changes in the solar core is long compared to
the $10^4$ yr  required for a photon to diffuse to the surface).

By combining the different experiments, each class of which has a different
spectral sensitivity, one concludes that
\beq
\frac{\phi(^7Be)}{\phi(^7Be)_{\rm SSM}} \ll \frac{\phi(^8B)}{\phi(^8B)_{\rm SSM}},
\eeq
where SSM refers to the standard solar model predictions. 
\begin{figure}[htb]
\vspace{9pt}
\postbb{18 183 557 555}{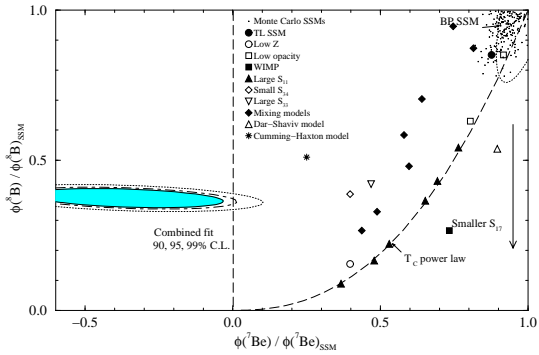}{0.95}
\caption{Allowed regions for the $^7Be$ and $^8B$ fluxes (normalized
by BP 98),
compared with  the predictions and uncertainties in the SSM
and various non-standard solar models. Courtesy of N. Hata.}
\label{BeB}
\end{figure}
The same
result holds even if one discards any one of the three types of experiment
(chlorine, gallium, water), {\it or}\ ignores the luminosity constraint.
No plausible astrophysical model has succeeded in suppressing $^7Be$
neutrinos significantly more than $^8B$ neutrinos, mainly because $^8B$ is made from
$^7Be$. Models with a lower core temperature or with a lower $S_{17}$ do
not come anywhere near the data. The
Cumming-Haxton model~\cite{cumming} with large
$^3He$ diffusion comes closest, but even that is far from the data.  That
model is probably also excluded by helioseismology, but Haxton has 
argued~\cite{haxton} that
final judgment should wait until a self-consistent model with $^3He$ diffusion
is constructed to be compared with the helioseismology data.

\subsection{Possible Solutions}
As discussed in the previous section, an astrophysical/nuclear
explanation of the solar neutrinos experiments is unlikely. The
most likely particle physics explanations include:
\begin{itemize}
 \item A matter enhanced (MSW) transition of $\nu_e$ into $\nu_\mu$ or
 $\nu_\tau$. There are the familiar small (SMA) and
large (LMA) mixing angle solutions~\cite{analyses}
 with $\Delta m^2 \sim
10^{-5}$ eV$^2$, as well as the low mass (LOW) solution with $\Delta m^2 \sim
10^{-7}$ eV$^2$  and near maximal mixing. The latter is a very poor fit, but
sometimes shows up in fits at the  99\% cl.
\begin{figure}[htb]
\vspace{9pt}
\postbb{73 85 517 654}{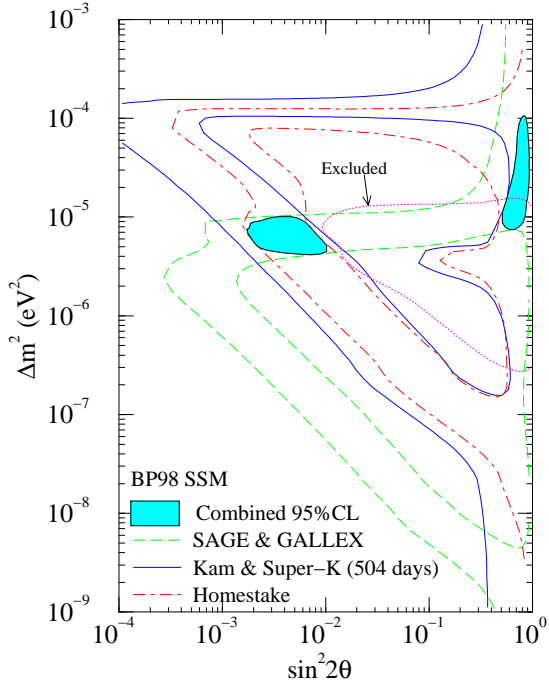}{0.95}
\caption{Allowed MSW solutions, not including Superkamiokande spectral data.
Courtesy of N. Hata.}
\label{mswregions}
\end{figure}
 \item There is also a small mixing angle MSW solution for $\nu_e$ into
a sterile neutrino $\nu_s$. The major difference between $\nu_{\mu,\tau}$ and
$\nu_s$, and the reason there is no LMA solution,
 is that in the first case the
$\nu_{\mu,\tau}$ can scattering elastically from electrons in the water Cerenkov
experiments, with about
$1/6^{th}$ the $\nu_e$ cross section, leading to a lower survival probability
for $\nu_e$ than for astrophysical or sterile neutrino solutions. There is
also a small difference for the MSW conversion rate for sterile neutrinos
in the Sun, but that is proportional to the neutron density, and is much less
important.
 \item The vacuum (``just so''~\cite{vacuum}) oscillation
solutions~\cite{analyses},
 with near maximal
mixing and $\Delta m^2 \sim 10^{-10}$
eV$^2$ are another possibility. These are somewhat fine-tuned, with
$\Delta m^2$ such that the Earth-Sun distance is at roughly half an oscillation
length, $L_{\rm osc}$,
 or an odd multiple. Since  $L_{\rm osc} = 4 \pi E/\Delta m^2$,
one expects a
significant variation of the $\nu_e$ survival  probability with neutrino energy.
 \item The above solutions are such that only two neutrinos are
important for the Solar neutrinos. However, it is possible that
 transitions between all three neutrinos are important.
There could be generalized  MSW solutions involving more than one
value of $\Delta m^2$, or mixed MSW and  vacuum solutions~\cite{hybrid}. In both
cases, there could be considerably different spectral distortions than
in the two-neutrino case. 
\end{itemize}

\noindent
Other possibilities include:
  \begin{itemize}
   \item Maximal mixing~\cite{maxmix}
   (i.e., vacuum oscillations with $\Delta m^2 \gg 10^{-10}$
eV$^2$), combined with a low $S_{17}$. Such solutions lead to an
energy independent suppression of the $\nu_e$ survival probability.
Even allowing a suppressed $^8B$ production rate, this possibility is viable
only if one ignores (or greatly expands the uncertainties in) the Homestake
Chlorine experiment.
   \item RSFP~\cite{RSFP}
   (resonant spin flavor precession), involving rotations of
left handed neutrinos into sterile right handed neutrinos, combined
with MSW flavor transitions. These were motivated by possible hints
(not confirmed by other experiments) of time dependence correlated with
the Sunspot activity in the chlorine experiment. This could only occur
if there are extremely large neutrino electric or magnetic dipole moments
or transition moments, which would present a considerable challenge to the
model builder. Although
such effects have not been reported by other groups, there is still a somewhat
surprising difference in rates observed by the GALLEX collaboration in their third and
fourth data taking intervals. However, this could also be a statistical fluctuation.
In any case, such RSFP effects could be probed experimentally by studying
the $\bar{\nu}_e$ and $\bar{\nu}_\mu$ spectra~\cite{RSFPspectra}.
   \item Flavor changing neutral current effects~\cite{FCNC}, possibly generated
by $R$-parity violating terms in supersymmetry, could be an alternative
means of generating enhanced neutrino flavor conversions in the Sun.
   \item The possible violation of Lorentz invariance~\cite{lorentz}
   could affect not
only the Solar neutrinos, but could also be relevant to the observed ultra
high energy cosmic rays.
   \item There could be a lepton flavor dependent violation of the the
equivalence principle~\cite{equiv}.
  \end{itemize}

 Perhaps the most important possibility or complication is that
more than one thing could be going on simultaneously. There
could be any of the above effects in conjunction with non-standard
properties of the Sun or nuclear cross sections. Many but not all such
NSSM possibilities are excluded by helioseismology and neutrino source experiments.
While it is very unlikely than such effects could by themselves account for the data,
their combination with new neutrino properties could considerably confuse the
interpretation of future experimental results. This is one or the reasons
that it is important to have as many independent precise experimental results
as possible.

\subsection{Needs}
To distinguish the many possibilities we need as much precise data
as possible. Especially useful are observables that are
independent of or insensitive to the initial $\nu_e$ fluxes, and therefore to the
astrophysical and nuclear cross section uncertainties. Such observables
include:

\begin{itemize}
\item The neutral to charged current interaction ratio (NC/CC), 
which will be measured by SNO for deuteron dissociation. 
Since the NC cross section is the same for all active
neutrinos, the NC rate measures the sum of the
$\nu_e$, $\nu_\mu$, and $\nu_\tau$ fluxes, while the CC
only measures $\nu_e$. An anomalous NC/CC ratio would provide
definitive evidence for transitions of $\nu_e$ into $\nu_\mu$ or $\nu_\tau$,
either by MSW or vacuum oscillations. Although the NC measurement is
difficult, SNO should have the requisite sensitivity. A confirmation
could be obtained by comparing the SNO CC rate with the fluxes determined
in $\nu e \RA \nu e$ measurements, since $\nu_{\mu,\tau}$ also contribute
to the latter, with about 1/6$^{\rm th}$ the $\nu_e$ cross section. 
(The Borexino experiment will similarly allow an indirect determination
of the transitions of $^7Be$ neutrinos into $\nu_{\mu,\tau}$ by comparing  with the
$\nu_e$ flux inferred from radiochemical experiments.) 

Transitions of $\nu_e$ into a sterile neutrino $\nu_s$ would  not lead to an
anomalous NC/CC ratio. This would make it much harder to verify $\nu_s$
transitions, but would serve as evidence for sterile neutrinos if 
MSW or vacuum oscillations are established by other means.

\item There is no known astrophysical mechanism that can significantly
distort the $^8B$ neutrino spectrum from the expected
$\beta$ decay shape. 
Not only would a spectral
distortion establish a non-astrophysical solution to the solar neutrinos,
but it would be a powerful probe of the mechanism. 
Study of the $^8B$ spectrum can be viewed as a cleaner extension
(by individual experiments) of the ``global'' spectral distortion
inferred from the combined experiments.

One expects significant spectral distortions
for the MSW SMA solution, for vacuum
oscillations, and for hybrid solutions, but not for the LMA solution. 
The ratio of observed to expected spectrum can be conveniently parametrized by
the first two moments~\cite{moments}, i.e., a linear approximation, for the SMA case,
while the other cases can exhibit more complicated shapes.
Measurement of the spectral distortion is very difficult, and requires
excellent energy calibrations and extending the measurement to as low
an energy as possible. Both SuperKamiokande and SNO have the capability
to measure a spectral distortion. SuperK has the advantage of higher statistics.
However, the $\nu$ energy is shared between the final electron and neutrino,
so any spectral distortion is partially washed out in the
observed $e^-$ spectrum.
SNO, on the other hand, has the advantage that the electron in
the CC reaction carries all of the neutrino energy (plus the known binding
energy), leading to a harder electron spectrum and an essentially
direct measurement of the $\nu$ spectrum.

One of the highlights of this conference was the preliminary new
statistics-limited
SuperKamiokande spectrum, from around 6.7 to 14.5 MeV,
obtained after a series of careful calibrations of their detector using
an electron Linac~\cite{superk}.  
The lower energy data are consistent with
no distortions, but there is evidence for a significant excess of events
in the three energy bins above 13 MeV. These data, for the first time,
give a statistically significant indication of a spectral distortion: the no
oscillation hypothesis (and also LMA solution) is disfavored at the 95-99\% CL level. 
The SMA MSW solution is also a very poor fit, although it is
allowed at 95\% CL. The best fit favors vacuum oscillations. The favored
$\Delta m^2 \sim 4 \x 10^{-10}$ eV$^2$ gives a much better fit to the
data than for the lower range
$\Delta m^2$ around $10^{-10}$ eV$^2$ found in recent global analyses of the 
total event rates. However, new studies based on BP 98 with its larger
$S_{17}$ allow a larger $\Delta m^2$, consistent with the spectral distortions.

An alternate interpretation of the high energy excess is that the
flux of hep neutrinos ($^3He + p \RA ^4He + e^+ + \nu_e$) has been
seriously underestimated. Their flux would have to be larger by a
factor of twenty or so from the usual estimates for them to contribute
significantly to
the excess, but it has been emphasized that there is no
direct experimental measure of or rigorous theoretical bound on the
cross section~\cite{hep}. The issue can be resolved by a careful study of
the energy range 14-18.8 MeV, above the endpoint of the $^8B$ spectrum.
(The highest energy SuperK bin is centered above this endpoint, but there
is a significant energy uncertainty.)

The SuperK spectrum has important implications, but it is still preliminary.
In additional to finalizing the analysis, additional lower energy points
are expected that should help clarify the situation.

\item 
For some regions of MSW parameters, one expects an asymmetry
between day and night event rates due to regeneration of $\nu_e$
at night as the converted neutrinos travel through the Earth~\cite{daynight}.
Superkamionde has binned their data for daytime and for a number
of different nighttime zenith angles
(i.e., different paths through the earth). They see no evidence
for a zenith angle dependence, and their overall day-night asymmetry is
\beq \frac{D-N}{D+N} = -0.023 \pm 0.020 \pm 0.014, \eeq
where $D$  ($N$) refers to day (night) rates and the first (second)
error is statistical (systematic). The absence of an effect excludes
a significant region of MSW parameter space independent of the details
of the solar model (and with only a small uncertainly from the Earth's
density profile). This excludes  the lower $\Delta m^2$ part of the
LMA solution, but has little impact on the SMA solution. (The 
 part of the SMA solution  with the largest
$\sin^2 2 \theta$ was expected to have a barely
observable day-night asymmetry, but the effect is predicted to be smaller
with the new BP 98 fluxes, which shift the SMA region to slightly smaller
$\sin^2 2 \theta$.)

Several authors have emphasized recently that the Earth effect is
signficantly enhanced for neutrinos passing through the core of the
Earth~\cite{parametric}.
(There is an analogous effect for atmospheric neutrinos.) 
 This parametric (or oscillation length)
resonance, in which the oscillation
length is comparable to the diameter of the core, was included
automatically in previous numerical studies, but not explicitly commented
on. It is larger for transitions into
$\nu_{\mu,\tau}$ than for $\nu_s$. Since relatively few of the
solar neutrinos pass through the core for the existing high latitude
detectors, it has been suggested that there should be a dedicated
experiment at low latitude~\cite{equator}.

\item
For vacuum  oscillations~\cite{analyses}, the Earth-Sun distance
is typically at a node of the oscillations. This is somewhat
fine-tuned, leading to the name ``just so''. Since the oscillation
length is
$4 \pi E/\Delta m^2$ there
is a strong energy dependence to the survival probability. 
 One 
also expects a strong seasonal variation, due to the eccentricity of the
Earth's orbit. However, the seasonal variation can be partially washed out
as one averages over energies, so one should ideally measure the
spectral shape binned with respect to the time of the year~\cite{newvac}.

\item
RSFP could lead to long term variations in the neutrino flux, e.g.,
correlated with Sunspots or Solar magnetic fields. Other changing magnetic
effects could conceivably alter the solar neutrinos in other ways, e.g.,
by changing the local density. Only the Homestake experiment has
seen any significant hint of a time variation, and that hint has been
considerably weakened by more recent Homestake data.
Nevertheless, it is conceivable that time dependent effects are energy
dependent, and therefore different experiments have different sensitivity.
They could also have been somewhat hidden in
the water Cerenkov experiments because of the neutral current. It would be useful
to run all of the experiments simultaneously through a solar cycle.

\item
RSFP~\cite{RSFP} could also lead to the production of $\bar{\nu}_e$ which can
be observed in the SNO detector by the delayed coincidence
of the $\gamma$ ray emitted by the capture of the neutron
from $\bar{\nu}_e p \RA e^+ n$.

\end{itemize}

\subsection{Outlook}
The model independent observables that can be measured by
SuperK and SNO for the $^8B$ neutrinos should go far towards
distinguishing the different possibilities. However, it will
be especially difficult to establish transitions into sterile neutrinos.
It will also be very difficult to sort out what is happening in
a three-flavor or hybrid scenario, such as MSW transitions combined with
non-standard solar physics. For these reasons, we would like to have
accurate information on spectral distortions, day-night effects (especially
for neutrinos passing through the Earth core),  NC/CC ratios, and absolute
fluxes arriving at the Earth for
the $^7Be$ and $pp$ neutrinos as well. There is a strong need for the
next generations of experiments.

A challenging but realistic goal is to simultaneously establish the
neutrino mechanism(s) (e.g., MSW SMA solution), determine the 
neutrino parameters, {\em and} study the Sun~\cite{futuremodind}.
Even with existing data,
if one assumed two-flavor MSW but allowed
an arbitrary solar core temperature $T_C$,
it was possible to simultaneously determine the MSW parameters (with larger
uncertainties than when the SSM  is assumed) and $T_C$, with the
result that $T_C = 0.99^{+0.02}_{-0.03}$ with respect to the SSM prediction
of 1.57 $\x 10^7$ K~\cite{hl}.
In the future, it should be possible to determine the
neutrino parameters and simultanously the $^8B$ and $^7Be$ fluxes, for comparison with
the SSM predictions.
It will also be possible to constrain density fluctuations in
the Sun~\cite{density},
which can smear out the MSW affects. However, recent estimates suggest
that such effects are negligible~\cite{newdensity}.

To fully exploit the future data, it will be important to carry out
global analyses of all of the observables in all of the experiments
(possibly incorporating helioseismology data as well). Global analyses
are  difficult because of difficulties with systematic errors.
However, they often contain more information than the individual
experiments, and allow uniform treatment of theoretical uncertainties.
For this purpose, it is important that each experiment publish all
of their data, such as double binning the data with respect to
energy and zenith angle, including full
systematics and correlations.

\section{ATMOSPHERIC NEUTRINOS}

Although the prediction for the 
absolute number of $\mu$ or $e$ produced by the
interactions  of neutrinos produced in cosmic ray
interactions in the atmosphere has a theoretical uncertainty of around
20\%, it is believe that the ratio $N(\mu)/N(e)$
can be predicted to within 5\%~\cite{atmflux}.
To zeroth approximation, the ratio is just
two, independent of the details of the cosmic ray flux or interactions, 
because each produced pion decays into two
$\nu_\mu$ and one $\nu_e$ (I am not distinguishing $\nu$ from $\bar{\nu}$),
and for energies large compared to $m_\mu$ the interaction cross sections
are the same.
Of course, the actual ratio  depends on the neutrino energies, and therefore on the
 details of the
hadronic energies, polarization of the intermediate muons from $\pi$
decay, etc. 

For years the ratio $R$ of observed $N(\mu)/N(e)$, normalized by
the predicted value, found in the water Cerenkov experiments
(Kamiokande, IMB, SuperKamiokande) has been around 0.6~\cite{atmdata}.
This has
recently been confirmed by the higher SuperK~\cite{superkatm} statistics
($R = 0.63(3)(5)$ for sub-GeV events and 0.65(5)(8) for multi-GeV), 
and independently by the iron calorimeter experiment at Soudan~\cite{soudan}
(0.58(11)(5)) and by Macro~\cite{macro} (0.53(15) for upward events and
0.71(21) for stopping or downgoing events). This depletion of
$\mu$ events suggests the possibility of $\nu_\mu$ oscillations
into $\nu_e$, $\nu_\tau$, or $\nu_s$, with near-maximal mixing
($\sin^2 2 \theta > 0.8$) and $\Delta m^2 \sim 10^{-3} -10^{-2}$ eV$^2$.

To confirm oscillations, more detailed information is needed.
Already, the CHOOZ~\cite{CHOOZ}
(France) reactor $\bar{\nu}_e$ disappearance experiment
excludes the $\nu_\mu \RA \nu_e$ interpretation of the atmospheric
neutrino anomaly for $\Delta m^2 > 10^{-3}$ eV$^2$. This should be
extended by the coming Palo Verde experiment~\cite{pverde},
and the planned KamLand~\cite{kamland}
experiment at Kamiokande (sensitive to many
nearby reactors) should extend the senstivity down to the MSW LA
solar neutrino range.

In the future~\cite{longbase},
there will also be accelerator long baseline
experiments for $\nu_\mu \RA \nu_{e,\tau}$ appearance, or 
$\nu_\mu$ disappearance (into $\nu_{e,\tau,s}$). The KEK to Kamiokande (K2K)
experiment will be sensitive
 to $\nu_\mu$ disappearance down
to
$\Delta m^2 \sim 5 \x 10^{-3}$ eV$^2$, while the Fermilab to Soudan
(MINOS) experiment will probe both appearance and disappearance
down to $10^{-3}$ eV$^2$. There
are also proposals for a CERN to Gran Sasso experiment (ICARUS, OPERA),
which would be sensitive to most of the parameter range suggested
by Superkamiokande. These experiments should be able to confirm or
refute the atmospheric neutrino oscillations, 
except\footnote{The long baseline experiments were
proposed when the earlier Kamiokande results suggested a somewhat
larger $\Delta m^2 \sim   10^{-2}$ eV$^2$.}
 possibly for
the smallest $\Delta m^2 \sim  10^{-3}$ eV$^2$.

Much more detailed information can be derived from the atmospheric
neutrino data itself, by searching for indications of the
$\sin^2 (1.27 \Delta m^2 L/E)$ dependence of the  transition
probability characteristic of neutrino oscillations. ($L$ is the distance
traveled and $E$ is the neutrino energy.) 
This can be studied by considering the zenith angle distribution
for fixed neutrino energy (in practice, the data is divided into sub-GeV and
mutli-GeV bins), or by up-down asymmetries $(U-D)/(U+D)$, where $U$ and $D$
are respectively the number of up and downgoing muons or
electrons~\cite{udasym}.  The data can also be plotted as a function of
$L/E$, but that is less direct since the full neutrino energy is not measured
on an event by event basis in the water Cerenkov experiments.

The Kamiokande collaboration observed an indication of oscillations in
their zenith angle distribution for contained events~\cite{kamatm}. 
However, the new Superkamiokande
zenith angle distributions for contained events
have much better statistics. They
strongly indicate a zenith angle distribution in muon events consistent
with oscillations, with an enhanced effect in the multi-GeV
sample, consistent with expectations. There is no anomaly or excess
in the electron events. This implies that $\nu_\mu$ is oscillating into
$\nu_\tau$ or possibly a sterile neutrino $\nu_s$, and not into
$\nu_e$. The latter result confirms the conclusions of  CHOOZ.
(Subdominant oscillations into $\nu_e$ in three-neutrino
schemes are still possible.) The SuperK events virtually
establish neutrino oscillations. Independent evidence is obtained by
the zenith angle distributions for upward through-going muon events
from SuperK, MACRO\footnote{The MACRO results~\cite{macro}
are not in very good
agreement with oscillations, but the oscillation hypothesis nevertheless
fits much better than the no-oscillation case.}, and very preliminary
results from SOUDAN.

Future atmospheric neutrino observations could possibly shed further
light on the question of whether $\nu_\mu$ is oscillating into
$\nu_\tau$ or into $\nu_s$, although they are all very difficult.
These include (a) subtle (e.g., parametric resonance)
effects on neutrinos propagating
through the Earth's core~\cite{coreatm},
which would affect $\nu_\mu \RA \nu_s$, but
not $\nu_\mu \RA \nu_\tau$ (because $\nu_\mu$ and $\nu_\tau$ have
the same neutral current interactions). In either scenario, 
secondary $\nu_\mu \RA \nu_e$ oscillations would also be
modified by Earth core effects.
(b) The NC/CC ratio, including its zenith angle distribution and
up-down asymmetry~\cite{ncatm}. The NC rate could in principle be measured
in $\nu N \RA \nu \pi^0 X$, although this is a very difficult
measurement. The preliminary SuperK result~\cite{superk}
 $R(\pi^0/e) =
0.93(7)(19)$ on the ratio or $\pi^0$ to $e$ events compared to expectations
slightly  favors $\nu_\mu \RA \nu_\tau$ but does not exclude $\nu_\mu \RA
\nu_s$. (c) Direct observation of events in which $\nu_\tau$ produces a $\tau$
would  establish $\nu_\mu \RA \nu_\tau$ oscillations~\cite{tauatm}.
However, this is extremely difficult.

There may also be significant three neutrino effects. For example,
even if the dominant transition for the atmospheric neutrinos involves
$\nu_\mu \RA \nu_\tau$, there could be important subdominant $\nu_e$
effects.

There have been several careful phenomenological analyses of the
atmospheric neutrino data in two neutrino and three neutrino
mixing schemes~\cite{atmstudy}. One important theoretical issue posed by the
atmospheric neutrinos, is why is there nearly maximal mixing
(i.e., $\sin^2 2 \theta \sim 1$), when most theoretical
schemes involving hierarchies of neutrino masses, as well
as the analogs in the quark mixing sector, yield small mixings.

\section{IMPLICATIONS FOR NEUTRINO MIXING}
\subsection{The Global Picture}
\label{global}
Various scenarios for the neutrino spectrum are possible, depending
on which of the experimental indications one accepts. The simplest
scheme, which accounts for the Solar (S) and Atmospheric (A) neutrino
results, is that there are just three light neutrinos, all active,
and that the mass eigenstates $\nu_i$ have masses in a hierarchy,
analogous to the quarks and charged leptons. In that case, the
atmospheric and solar neutrino mass-squared differences are measures
of the mass-squares of the two heavier states, so that
$m_3 \sim (\Delta m^2_{\rm atm})^{1/2} \sim 0.03-0.1$ eV;
$m_2 \sim (\Delta m^2_{\rm solar})^{1/2} \sim 0.003$ eV (for MSW)
or $\sim 10^{-5}$ eV (vacuum oscillations), and $m_1 \ll m_2$.
The weak eigenstate neutrinos
$\nu_a =(\nu_e, \nu_\mu, \nu_\tau)$ are related to the mass eigenstates 
$\nu_i$ by a unitary transformation $\nu_a = U_{ai} \nu_i$.
If one makes the simplest assumption (from the Superkamiokande and CHOOZ data),
that the $\nu_e$ decouples entirely from the atmospheric neutrino oscillations,
$U_{e3}=0$,
(of course, one can relax this assumption somewhat) and ignores possible 
CP-violating phases~\cite{cp}, then 
\beqa
\left(  
\begin{array}{c}
\nu_e \\ \nu_\mu \\ \nu_\tau
\end{array}
\right)
& = &
\left(  
\begin{array}{ccc}
1 & 0 & 0 \\
0 & c_\alpha & -s_\alpha  \\
0 & s_\alpha  & c_\alpha
\end{array}
\right)  \nonumber \\
 & \x &
\left(  
\begin{array}{ccc}
c_\theta & -s_\theta & 0 \\ 
s_\theta & c_\theta & 0 \\
0 & 0 & 1
\end{array}
\right)
\left(  
\begin{array}{c}
\nu_1 \\ \nu_2 \\ \nu_3
\end{array}
\right),
\eeqa
where $\alpha$ and $\theta$ are mixing angles associated with the atmospheric
and solar neutrino oscillations, respectively, and where $c_\alpha \equiv 
\cos \alpha$, $s_\alpha \equiv \sin \alpha$, and similarly for $c_\theta, s_\theta$.

For maximal atmospheric neutrino mixing, $\sin^2 2 \alpha \sim 1$, this implies
$c_\alpha = s_\alpha = 1/\sqrt{2}$, so that
\beq
U = 
\left(  
\begin{array}{ccc}
c_\theta & -s_\theta & 0 \\ 
\frac{s_\theta}{\sqrt{2}} & \frac{c_\theta}{\sqrt{2}} & -\frac{1}{\sqrt{2}} \\
\frac{s_\theta}{\sqrt{2}} & \frac{c_\theta}{\sqrt{2}} & \frac{1}{\sqrt{2}}
\end{array}
\right).
\eeq
For small $\theta$, this implies that $\nu_{3,2} \sim \nu_{+,-} \equiv (\nu_\tau \pm
\nu_\mu)/\sqrt{2}$ participate in atmospheric oscillations, 
while the solar neutrinos
are associated with a small additional mixing between $\nu_e$ and $\nu_-$.
Another limit, suggested by the possibility of vacuum oscillations for
the solar neutrinos, is $\sin^2 2 \theta \sim 1$, or
$c_\theta = s_\theta = 1/\sqrt{2}$, yielding
\beq
U = 
\left(  
\begin{array}{ccc}
\frac{1}{\sqrt{2}} & -\frac{1}{\sqrt{2}} & 0 \\ 
\frac{1}{2} & \frac{1}{2} & -\frac{1}{\sqrt{2}} \\
\frac{1}{2} & \frac{1}{2} & \frac{1}{\sqrt{2}}
\end{array}
\right),
\eeq
which is referred to as  bi-maximal mixing~\cite{bimax}.
A number of authors have discussed this pattern and how it might
be obtained from models, as well as how
much freedom there is to relax the assumptions of maximal
atmospheric and solar mixing (the data actually allow $\sin^2 2 \alpha \simgr
0.8$ and $\sin^2 2 \theta \simgr 0.6$) or the complete decoupling of $\nu_e$
from the atmospheric neutrinos. Another popular pattern,
\beq
U = 
\left(  
\begin{array}{ccc}
\frac{1}{\sqrt{2}} & -\frac{1}{\sqrt{2}} & 0 \\ 
\frac{1}{\sqrt{6}} & \frac{1}{\sqrt{6}} & -\frac{2}{\sqrt{6}} \\
\frac{1}{\sqrt{3}} & \frac{1}{\sqrt{3}} & \frac{1}{\sqrt{3}}
\end{array}
\right),
\eeq
known  as democratic mixing~\cite{democratic}, yields maximal solar oscillations
and near-maximal ($8/9$) atmospheric oscillations.

In this hierarchical pattern, the masses are all too small
to be relevant to mixed dark matter (in which one of the components of
the dark matter is hot, i.e., massive neutrinos) 
or to neutrinoless double beta decay ($\beta \beta_{0\nu}$). However,
the solar and atmospheric oscillations only determine the differences in
mass squares, so a variant on this scenario is that the three
mass eigenstates are nearly degenerate rather than
hierarchical~\cite{degenerate},
with small splittings associated with $\Delta m^2_{\rm atm}$
and $\Delta m^2_{\rm solar}$. For the common mass $m_{\rm av}$
in the 1-several eV range, the hot dark matter could account for
the dark matter on large scales (with another, larger, component of cold
dark matter accounting for smaller structures)~\cite{mdm}. If the neutrinos
are Majorana they could also lead to $\beta \beta_{0\nu}$~\cite{nulessbb}.
Current limits imply an upper limit of
\beq \langle m_{\nu_e} \rangle = \sum_i \eta_i U_{ei}^2 |m_i|
< 0.46 -1\ {\rm eV}, \label{mnueff} \eeq
on the effective mass for a mixture of light Majorana mass eigenstates,
where $\eta_i$ is the CP-parity of $\nu_i$ and the
uncertainty on the right is due to the nuclear matrix elements.
(There is no constraint on Dirac neutrinos.)
The combination of small $\langle m_{\nu_e} \rangle \ll m_{\rm av}$,
maximal
atmospheric mixing, and $U_{e3}=0$ would imply cancellations, 
so that $\eta_1 \eta_2 =
-1$ and $c_\theta = s_\theta = 1/\sqrt{2}$, i.e., maximal solar mixing.
Even the more stringent limit in (\ref{mnueff}) is large
enough that there is room to relax all of these assumptions considerably.
Nevertheless, there is strong motivation to try to improve the
$\beta \beta_{0\nu}$ limits.

The LSND experiment~\cite{LSND} has reported evidence for $\nu_\mu \RA \nu_e$
and $\bar{\nu}_\mu \RA \bar{\nu}_e$ oscillations with
$\Delta m^2_{\rm LSND} \sim 1$ eV$^2$ and small mixing
$\sim 10^{-3}-10^{-2}$, while the KARMEN experiment sees no
candidates. KARMEN~\cite{KARMEN} is sensitive to most of the same parameter
range as LSND, although there is a small window of oscillation
parameters for which both experiments are consistent. A resolution
of the situation may have to wait for the mini-BOONE experiment at
Fermilab. However, it is interesting to consider the implications
if the LSND result is confirmed. In that case, there
are three distinct mass-squared differences,
$\Delta m^2_{\rm LSND} \sim 1$ eV$^2$,
$\Delta m^2_{\rm atm} \sim 10^{-3}-10^{-2}$ eV$^2$, and
$\Delta m^2_{\rm solar} \sim 10^{-5}$ eV$^2$ (MSW) or $10^{-10}$ eV$^2$
(vacuum), implying the need for a fourth neutrino\footnote{There have
been several attempts to get by with only three neutrinos. However,
attempts to take $\Delta m^2_{\rm solar} =
\Delta m^2_{\rm atm}$~\cite{threenua}
fail because they lead to an unacceptable energy-independent suppression of 
the solar neutrinos. Similarly, $\Delta m^2_{\rm atm} = \Delta m^2_{\rm LSND}$
were marginally compatible with the earlier Kamiokande atmospheric
data~\cite{threenub},
but do not describe the zenith angle distortions (and lower
$\Delta m^2_{\rm atm}$) observed by Superkamiokande. There is still
a possibility of combining a three neutrino scheme with anomalous
interactions~\cite{anomalous}, which could, e.g., affect the zenith
distribution and  allow a larger $\Delta m^2_{\rm atm}$, or affect the LSND
results and allow a lower $\Delta m^2_{\rm LSND}$.}.
Since the $Z$ lineshape measurements at LEP only allow $2.992 \pm 0.011$
light, active neutrinos~\cite{lineshape},
any light fourth neutrino would have to be sterile, 
$\nu_s$.

Several mass patterns for the four neutrinos have been suggested~\cite{fournu}.
(There course also be more than four light neutrinos~\cite{sixnu}.)
To be consistent with both LSND and CHOOZ,  states containing
$\nu_\mu$ and $\nu_e$ must be separated by about 1 eV.
Assuming the atmospheric neutrinos involve $\nu_\mu \RA \nu_\tau$,
one could have nearly degenerate $\nu_{+,-} \equiv (\nu_\tau \pm
\nu_\mu)/\sqrt{2}$ at around 1 eV, with the solar neutrinos described
by a dominantly
$\nu_s$ state at  $\sim 0.003$ eV 
or $\sim 10^{-5}$ eV and a much lighter (dominantly) $\nu_e$.
(Solar neutrinos can be accounted for by a SMA MSW solution
or possibly by vacuum oscillations, but not by a LMA MSW.)
Alternatively, one could reverse the pairing, with a
nearly degenerate $\nu_s$ and $\nu_e$ at $\sim$ 1 eV,
and $\nu_{+,-}$ around 0.03-0.1 eV.
The other models involve $\nu_\mu \RA \nu_s$ with near-maximal
mixing for
the atmospheric neutrinos, and $\nu_e \RA \nu_\tau$
for the solar neutrinos. Again, there are two
possibilities, with the nearly degenerate $\nu_s-\nu_\tau$ pair
around 1 eV and a lighter $\nu_\tau-\nu_e$, or the other way
around. 

All of these patterns involve two neutrinos in the eV
range, and therefore the possibility of a significant hot dark
matter component. The two which have the (dominantly) $\nu_e$
state around 1 eV could contribute to $\beta \beta_{0\nu}$
if the neutrinos are Majorana. A very
small $\langle m_{\nu_e} \rangle$ due to cancellations would
suggest near maximal mixing for the solar neutrinos, but
this could again be relaxed significantly given all of the
uncertainties.

\section{PARTICLE PHYSICS IMPLICATIONS: FROM THE TOP DOWN}
Almost all extensions of the standard model predict non-zero neutrino
mass at some level, often in the observable $10^{-5}-10$ eV range.
It is therefore difficult to infer the underlying physics from
the observed neutrino masses. However, the neutrino mass spectrum
should be extremely useful for top-down physics; i.e.,  the
predicted neutrino masses and mixings should provide an
important test, complementary to, e.g., the sparticle, Higgs,
and ordinary fermion spectrum, of any
concrete fundamental theory with serious predictive power.

Prior to the precision $Z$-pole measurements at LEP and SLC
there were two promising paths for physics beyond the standard
model:  compositeness at the TeV scale (e.g., dynamical symmetry
breaking, composite Higgs, or composite fermions), or unification,
which most likely would have led to deviations from the
standard model prediction at the few \% or few tenths of a \% level, 
respectively. The absence of large deviations~\cite{precision}
strongly supports
the unification route, which is the domain of supersymmetry,
grand unification, and superstring theory. The implication is that
non-zero neutrino masses are most likely not the result of
unexpected new
physics at the TeV scale, such as by  loop effects
associated with new ad hoc scalar fields. (They could, however, be
due to neutrino-neutralino mixing or loop effects  
in supersymmetric
models with $R$ parity breaking.) Alternatively, they
could be associated with new physics at very high energy scales,
most likely either
seesaw models  or  higher
dimensional operators.

\section{ORDINARY-STERILE NEUTRINO MIXING}

As discussed in Section \ref{global} the combination of solar
neutrinos, atmospheric neutrino oscillations, and the LSND results,
if confirmed, would most likely imply the mixing of ordinary
active neutrinos with one (or more) light sterile neutrinos.
One difficulty is that the sterile neutrinos could have been
produced in the early universe by the mixings. For the range
of mass differences and mixings relevant to LSND and the
atmospheric neutrinos, the sterile neutrino would have been produced
prior to nucleosynthesis, changing the freezeout temperature
for $\nu_e n \leftrightarrow e^- p$ and leading to too much $^4He$~\cite{bbn}.
However, Foot and Volkas have recently~\cite{fv}
argued that MSW effects involving
sterile neutrinos could amplify a small lepton asymmetry, leading to an excess
of $\nu_e$ compared to $\bar{\nu}_e$, reducing the $^4He$.
It has also been argued that
ordinary-sterile neutrino mixing could facilitate
heavy element synthesis by $r$-processes
in the ejecta of
neutrino-heated supernova explosions~\cite{rprocess,bbn}. 

Most extensions of the standard model predict the existence of sterile
neutrinos. For example, simple $SO(10)$ and $E_6$ grand unified theories
predict one or two sterile neutrinos per family, respectively. The only real
questions are whether the ordinary and sterile neutrinos of the same chirality mix
significantly with each other, and whether the mass eigenstate neutrinos are
sufficiently light. When there are only Dirac masses, the ordinary and sterile
states do not mix because of the conserved lepton number. Pure
Majorana masses do not mix the ordinary and sterile sectors either.
In the seesaw model the mixing is negligibly small, and the 
(mainly) sterile eigenstates are too heavy to be relevant to oscillations.
The only way to have significant mixing and  small mass eigenstates
is for the Dirac and Majorana neutrino mass terms to be extremely
small and to also be comparable to each
other.
This appears to
require two miracles in conventional models of neutrino mass.

One promising possibility involves the generation of neutrino
masses from higher-dimensional operators in theories involving
an intermediate scale~\cite{nro},
as described in Section \ref{theorysection}.
Depending on the intermediate scale and the dimensions of the operators
naturally small Dirac and Majorana masses are possible, and in
some cases they are automatically of the same order of
magnitude~\cite{sterile}.
Another interesting possibility~\cite{parallel} involves sterile neutrinos
associated with a parallel hidden sector of nature as suggested in
some superstring and supergravity theories. Other mechanisms
in which one can obtain ordinary-sterile neutrino mixing are
described in~\cite{othersterile}.

\section{CONCLUSIONS}

\begin{itemize}
\item Neutrino mass is an important probe of particle physics, astrophysics,
and cosmology.
\item There are several experimental indications or suggestions:
 (a) The Superkamiokande and other results on atmospheric
   neutrinos provide strong evidence for $\nu_\mu$ oscillations.
   (b) The combination of solar neutrino experiments implies a global
   spectral distortion, strongly supporting neutrino transitions or
   oscillations. The preliminary SuperK results on the $^8B$ spectrum
   suggests a spectral distortion, most consistent with vacuum oscillations
   but possibly with small angle MSW.
   (c) LSND has candidate events in both decay at rest and decay in
   flight. The non-observation of candidates by KARMEN is close to being
   an experimental contradiction, also there is still a small parameter 
   space consistent with both.
   (d) Mixed dark matter is an interesting hint for eV scale masses,
   but is not established.
\item In the future many solar neutrino 
   experiments and (model independent) observables will be
   needed to identify the mechanism, determine the neutrino parameters,
  and simultaneously study the Sun. This program is complicated by
  possible three neutrino effects, possible sterile neutrinos, and
  the possibility that there are both neutrino mass effects and
  nonstandard solar physics (although the latter is constrained by
  helioseismology). Experiments that are sensitive to the $pp$ and $^7Be$
  neutrinos are needed. Important observables include neutral to charged
  current ratios, spectral distortions, day-night effects (possibly
  involving parametric core enhancement), and seasonal variations
  (especially for vacuum oscillations).
  \item For the atmospheric neutrinos, we need more detailed spectral and
  zenith angle information, and the neutral to charged current ratio as a
  function of the zenith angle. Independent information, including possible
  $\nu_\tau$ appearance, for the
   same parameter range should be forthcoming from long baseline experiments.  
\item The planned Mini-BOONE experiment at Fermilab should clarify the
 LSND-KARMEN situation.
\item Future cosmic microwave anisotropy experiments and large scale
 sky surveys should be able to determine whether neutrinos contribute
 significantly to the dark matter.
\item Significant improvements in $\beta \beta_{0\nu}$ would be very
 powerful probes of the Majorana nature of neutrinos in the mass ranges
 suggested by the LSND and atmospheric neutrino results.
\item Most extensions of the standard model predict nonzero neutrino masses,
 so it is difficult to determine their origin in a ``bottom-up'' matter.
 However, the neutrino spectrum will be a powerful constraint on ``top-down''
 calculations of fundamental models.
\item The possibility of mixing between ordinary and light sterile neutrinos
 should be taken seriously.
\end{itemize}

\section*{ACKNOWLEDGMENTS} 
This work was supported by U.S. Department of Energy Grant No. DOE-EY-76-02-3071.
I am grateful to Naoya Hata for collaborations on the implications
of Solar neutrinos. 



\begin{thebibliography}{99}
\bibitem{atmdata}
References to earlier work may be found in~\cite{superkatm}.
\end{thebibliography}
\end{document}